**Kinetic roughening transition of ice crystal and its implication during recrystallization**

Jorge H. Melillo[1] and Ido Braslavsky[1]

[1]Institute of Biochemistry, Food Science, and Nutrition, Robert H. Smith Faculty of Agriculture, Food and Environment, The Hebrew University of Jerusalem, Rehovot 7610001, Israel

Corresponding author: Ido Braslavsky
Mail: ido.braslavsky@mail.huji.ac.il

**Graphical Abstract**

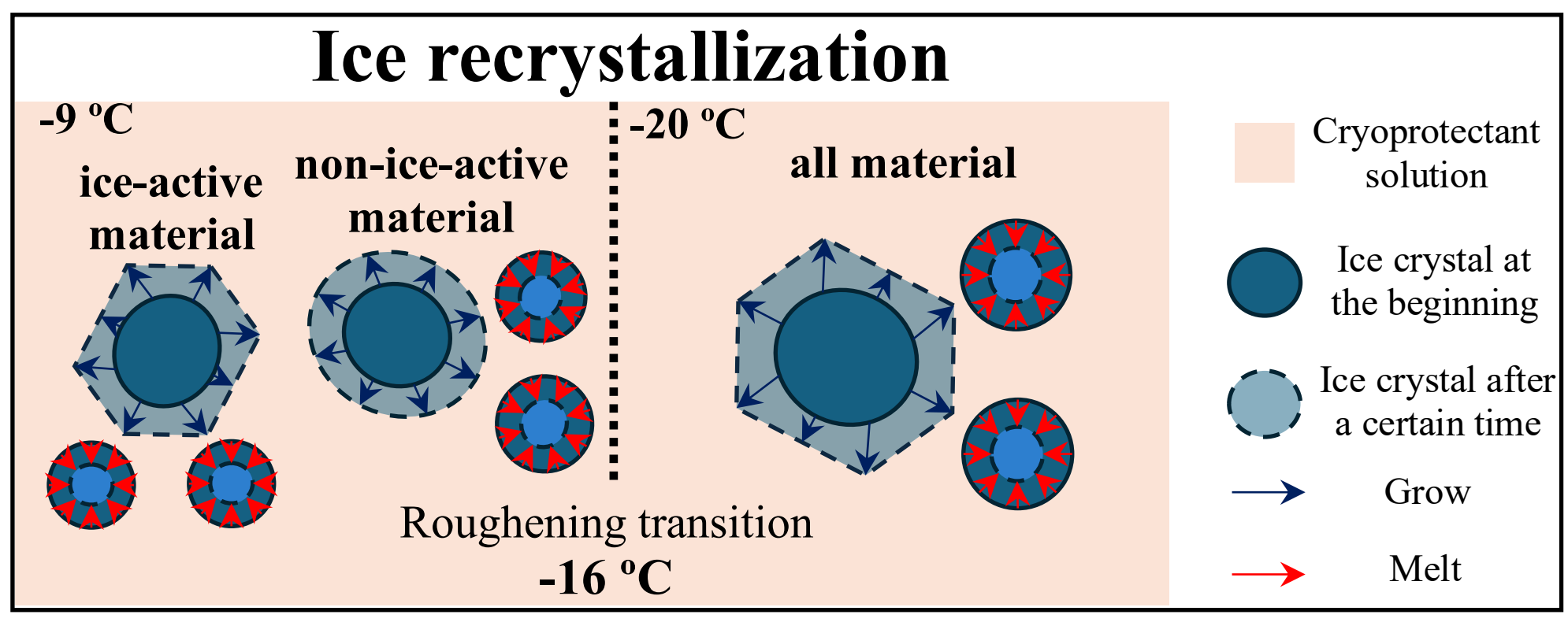

**Abstract**

*Hypothesis*
Roughening transitions at solid–liquid interfaces govern crystal morphology in diverse systems. In ice crystallization, these transitions control interfacial faceting and surface kinetics. Faceted morphologies are often associated with ice-active molecules, which inhibit recrystallization and are essential for cryopreservation. We hypothesize that kinetic roughening transitions can induce faceting even in the absence of ice-active agents, particularly at high solute concentrations with depressed melting points, potentially complicating the interpretation of crystal morphology as an indicator of ice activity.

*Experiments*
We investigated the kinetic roughening transition of ice in dimethyl sulfoxide (DMSO) and proline–water solutions using cryomicroscopy and real-time image analysis. Crystals grew in microdroplets, maintaining near-equilibrium conditions as solute concentration increased during growth due to conversion of liquid water to ice. Antifreeze protein type III (AFPIII) was applied to distinguish intrinsic roughening from adsorption-mediated effects.

*Findings*
A distinct kinetic roughening transition temperature ($T_R$ = -16.0 ± 0.2 °C) was identified, marking a shift from rounded disks at higher temperatures to faceted hexagonal plates at lower temperatures, independent of solute type. Recrystallization below $T_R$ revealed asymmetry between growth and melting interfaces. AFPIII promoted faceting even above $T_R$, consistent with stabilization of step edges and elevation of the roughening transition temperature. These results clarify the interplay between intrinsic interface kinetics and molecular adsorption, with implications for interpreting ice morphology, surface roughening, and cryopreservation design.

## 1. Introduction

Ice growth is one of the most visually captivating phenomena in science. The Nakaya diagram, for example, elegantly illustrates the diverse shapes of snow crystals formed under varying atmospheric conditions[1]. These patterns are replicated in the laboratory by adjusting temperature, pressure, and humidity, yielding precise crystal shapes. For instance, in vapour, near -2 ºC, plate-like growth predominates: thick plates form at low supersaturation, thinner plates at intermediate supersaturation, and dendritic plates at high supersaturation. At temperatures near -5 ºC, columnar growth emerges, producing stout columns at low supersaturation, slender hollow columns at intermediate supersaturation, and clusters of needle-like crystals at higher supersaturation[1,2]. Such diversity in morphology originates from the anisotropy in surface kinetics across crystallographic planes.

While vapor-grown ice has been extensively studied, ice grown from the liquid phase exhibits fundamentally different behaviour. In these systems, crystals develop solutions cooled below their equilibrium melting point, where supercooling is defined as the difference between the bath temperature and the equilibrium temperature. Under mild supercooling conditions, when the system remains near equilibrium, ice crystals tend to grow as circular disks, displaying a faceted basal plane and a rough prism plane. As supercooling increases, the morphology becomes more complex: first, hexagonal dendritic structures emerge, followed by double-pyramidal dendrites when supercooling exceeds approximately 2.7 °C[3,4]. These transitions reflect the strong dependence of crystal shape on interfacial kinetics. To avoid these rapid, non-equilibrium growth modes, the system's melting point must be depressed, either by increasing pressure or by introducing solutes[5] such as dimethyl sulfoxide (DMSO), fructose[6], and glycerol. Those cryoprotectants suppress freezing through colligative effects that lower the chemical potential of water, rather than by direct interaction with ice[7]. This controlled suppression enables the study of intrinsic surface kinetics under stable, slow-growth conditions.

Crystal shape can also depend on the kinetic roughening transition (KR), a temperature-dependent shift in surface morphology during slow crystal growth. Below the KR temperature ($T_R$), certain crystal faces grow near equilibrium via the propagation of molecular layers, resulting in a faceted appearance. Under these conditions, the formation of new layers typically requires crystal defects to nucleate[8]. Above $T_R$, at growth near equilibrium conditions, molecular steps can nucleate spontaneously without an energy barrier [9]. As a result, the surface becomes microscopically rough and loses its ability to maintain a flat, stepwise growth front. This leads to a rounded or curved macroscopic crystal shape. The transition reflects a change in the molecular attachment mechanism and affects both the growth dynamics and final morphology[9]. Vapor-grown ice crystals exhibit KR, with prism faces becoming rough above approximately -2 °C, while basal faces remain faceted up to the melting point. At the ice-liquid interface, the prism faces undergo a KR at -16 ºC, as measured under high-pressure conditions near equilibrium[10–12]. Importantly, the pressure conditions relevant for KR, less than 2000 bar, are lower than those at which new high-pressure ice phases appear[13–15]. Above $T_R$, ice crystals in water or solutions typically form circular disks bounded by basal facets and rounded prism faces. Below -16 °C, these disks transform into hexagonal plates with well-defined prism facets. Recrystallization experiments in high concentrations of fructose exhibit faceted hexagonal crystals[6], suggesting a KR occurs due to the depressed melting point at high solute concentration, although the transition itself was not directly measured in those experiments.

However, similar hexagonal morphologies can also arise in the presence of ice-active agents such as antifreeze proteins (AFPs), which modulate crystal shape through molecular adsorption[16–19]. There are also some salts, such as NaF and $NH_4Cl$, that can change the morphology of the ice crystals above $T_R$[20]. This resemblance complicates efforts to distinguish intrinsic roughening from adsorptive effects. AFPs typically bind to specific crystallographic planes, inhibiting ice growth and promoting faceted structures[19,21–24]. Therefore, decoupling the influence of temperature-dependent interface kinetics from that of solute adsorption is essential for accurate interpretation of ice morphologies in solution systems.

Lowering the melting temperature is also central to cryopreservation strategies, where solutes such as DMSO and glycerol act as cryoprotectant by vitrifying the surrounding solution and stabilizing biomolecules. These solutes depress the melting point through entropy-driven colligative effects rather than direct interaction with the ice interface[25]. In contrast, AFPs exert their effects via adsorption to specific planes, influencing crystal growth direction and shape. AFP efficacy is typically evaluated through thermal hysteresis, the gap between melting and freezing temperatures[26]. Yet even AFPs with negligible thermal hysteresis can display strong ice recrystallization inhibition[24,27] (IRI), where larger crystals grow at the expense of smaller ones.

The mechanism by which AFPs function is through direct binding to ice, which is why they are termed ice-binding proteins[19,28]. This interaction is mediated by ice-binding sites that stabilize the adjacent water network in a configuration complementary to the ice lattice[23]. Large ice-binding proteins, such as ice nucleating proteins, can promote ice formation by acting as nucleation enhancers as they stabilize large ice nucleai[29,30]. In contrast, typical AFPs are relatively small and generally do not act as strong nucleators, although some studies suggest they may exhibit weak nucleation activity under specific conditions[21,31,32]. Their principal role is to bind and “pin down” the growing ice surface, a process known as adsorption inhibition, first proposed by Raymond and DeVries[33].

Notably, certain non-protein solutes, such as polyvinyl alcohol[34], zirconium acetate[35], and nanocellulose[36], also exhibit IRI activity in the absence of significant thermal hysteresis.

Melting can alter the morphology of faceted ice crystals in characteristic ways. During melting, hexagonally faceted ice crystals tend to lose their faceted shape. Crystal corners melt first and facets persist longer and the shape changes to a non-faceted hexagonal[12]. This phenomenon was also recorded in ice crystals in the presence of hyperactive AFPs in two dimensions[37] as well in three dimensions[16,38]. Upon cooling slightly below the melting point, the ice regrows into a faceted hexagonal configuration identical to the initial state but rotated relative to the melting structure[12,16,37,38]. These observations underscore that melting and growth may follow distinct morphological pathways, influenced by both thermodynamic and interfacial kinetic factors.

As mentioned above, numerous studies have observed that the basal face of ice remains faceted up to the melting point, lacking the roughness exhibited by the prism face. Several factors may relate to this difference. For instance, the basal face grows slower than the prism face[39,40]. Jackson *et al*[41] noted that the growth along the basal face (c-axis) requires the cooperative attachment of at least four water molecules forming a six-bond cluster, a configuration that is stable but seldom forms spontaneously. In contrast, the prism face (a-axis) requires only a two-molecule cluster stabilized by three hydrogen bonds. Although both arrangements achieve a similar average bond-per-molecule ratio, the probability for step nucleation differ substantially, which result in difference in roughening transition temperature between the two facets.

Thus, while the basal face grows in an ordered bilayer-by-bilayer mechanism, the prism face exhibits a collected molecule process[42,43]. Despite these observations, the question of why the basal face remains faceted while the prism face becomes rough remains unresolved. This same question was recently revisited by Mochizuki et al.[44], who were unable to provide a definitive answer. Based on Mukherjee and Bagchi study[45], Mochizuki et al. suggested that entropy, specifically the rotational entropy of water molecules, could play a key role in maintaining the well-defined structure of the basal face[44,45]. These studies suggest that the basal face resists roughening due to kinetic and entropic barriers, while the prism face, with faster growth and lower entropic resistance, undergoes a KR.

In this study, we explored the KR of ice grown from DMSO and proline–water solutions under atmospheric pressure. These solutes depress the melting point without binding to ice, providing a clean system to study intrinsic faceting behaviour, as mentioned above. Using cryomicroscopy and real-time image analysis, we identify a distinct roughening transition temperature ($T_R$ = -16.0 ± 0.2 °C), where ice morphology shifts from rounded disks to hexagonal plates, independent of solute type, and matching values reported only under high-pressure conditions. We further reveal that while ice growth across $T_R$ follows a sigmoidal transition, melting proceeds via exponential relaxation, demonstrating a previously unreported asymmetry in interface kinetics. Finally, we show that even sub-functional concentrations of antifreeze protein type III (AFPIII) modulate faceting above $T_R$, elevating the roughening threshold and illustrating how molecular adsorption controls surface roughness transitions.

These findings distinguish true KR from faceted morphologies induced by ice-active molecules. While both can produce similar crystal shapes, they arise from fundamentally different mechanisms: intrinsic interface kinetics versus adsorption-mediated shaping. Recognizing this distinction is essential for correctly interpreting ice morphology in systems involving cryoprotectants or antifreeze agents. This clarification has broad implications for cryobiology, interfacial science, and atmospheric processes, where morphological features are often used to infer molecular activity.

## 2. Materials and Methods

### 2.1 Sample Preparation

Dimethyl sulfoxide (DMSO, ≥99.9%, CAS No. 67-68-5, Sigma-Aldrich, D4540) and L-proline (≥99.9%, CAS No. 147-85-3, Sigma-Aldrich, P5607) were purchased from Sigma-Aldrich (St. Louis, MO, USA) and used without further purification. Ultrapure water (resistivity ≥ 18.2 MΩ·cm; conductivity ≤ 0.055 µS/cm) was obtained from a Barnstead™ GenPure™ xCAD Plus water purification system (Thermo Scientific, Waltham, MA, USA) and was added to achieve the appropriate water concentration for preparing the aqueous solutions. DMSO solutions were prepared at concentrations of 10, 20, 25, 30 and 40 % (v/v), while proline solutions were prepared t concentrations of 20, 30 and 40 % wt. Table 1 presents the melting temperatures of DMSO and proline solutions as estimated from our experimental observations. For comparison, values for DMSO solutions obtained from the literature[46], are also shown in parentheses.

**Table 1.** Melting temperatures of DMSO and proline aqueous solutions at various concentrations.

| | DMSO [% (v/v)] | | | | | Proline [% wt] | | |
|---|---|---|---|---|---|---|---|---|
| | 10 | 20 | 25 | 30 | 40 | 20 | 30 | 40 |
| Melting Temperature [°C] | -3.5 (-3.2) | -8.7 (-8.0) | -10.6 (-11.3) | -17.7 (-15.6) | -31.0 (-29.0) | -6.5 | -10.5 | -16.4 |

AFPIII was expressed in *E. coli* BL21-DE3-PlysS and purified as described by Sirotinskaya et al[47]. Briefly, bacteria were grown in Luria Broth agar and Terrific Broth media with ampicillin at 37 °C, harvested by centrifugation, and stored at -80 °C. Cells were lysed by sonication, and the protein was purified using Ni-NTA affinity chromatography or the falling water ice purification method[48,49]. Purity and activity were verified by SDS-PAGE and thermal hysteresis assays, respectively. The purified AFPIII was first diluted in pure water to final concentrations of 0.1, 0.5, and 1 μM. DMSO was then added to each solution to achieve a final composition of 20% (v/v) DMSO.

For single-crystal analysis, emulsions of the solutions with Immersion Oil Type B (analytical grade, Electron Microscopy Sciences, Hatfield, PA, USA; Cat. No. 12608-01), were prepared. For the emulsions, we used a ratio of approximately 5% solution to 95% oil and mixed with the tip of a pipette for around 10 minutes.

2.2 Cryomicroscopy

KR experiment was conducted using a Linkam MDBCS196 cold stage (Linkam Scientific Instruments Ltd., Tadworth, UK) mounted on an Olympus BX41 microscope (Olympus Corporation, Tokyo, Japan). The system was controlled by a T95-Linkam controller equipped with an LNP95 liquid nitrogen cooling pump (Linkam Scientific Instruments Ltd., Tadworth, UK). For single crystal experiments, the emulsion was placed on a 16 mm glass disk, and measurements were conducted with another 10 mm glass disk cover. The quantity of the emulsion was not measured. For recrystallization experiments, 1.6 μL of the sample was placed on a 20 mm sapphire disk and covered with a 10 mm diameter glass disk. The sample was sealed with immersion oil to prevent evaporation. Each experiments were calibrated with the melting temperature of ultra pure water.

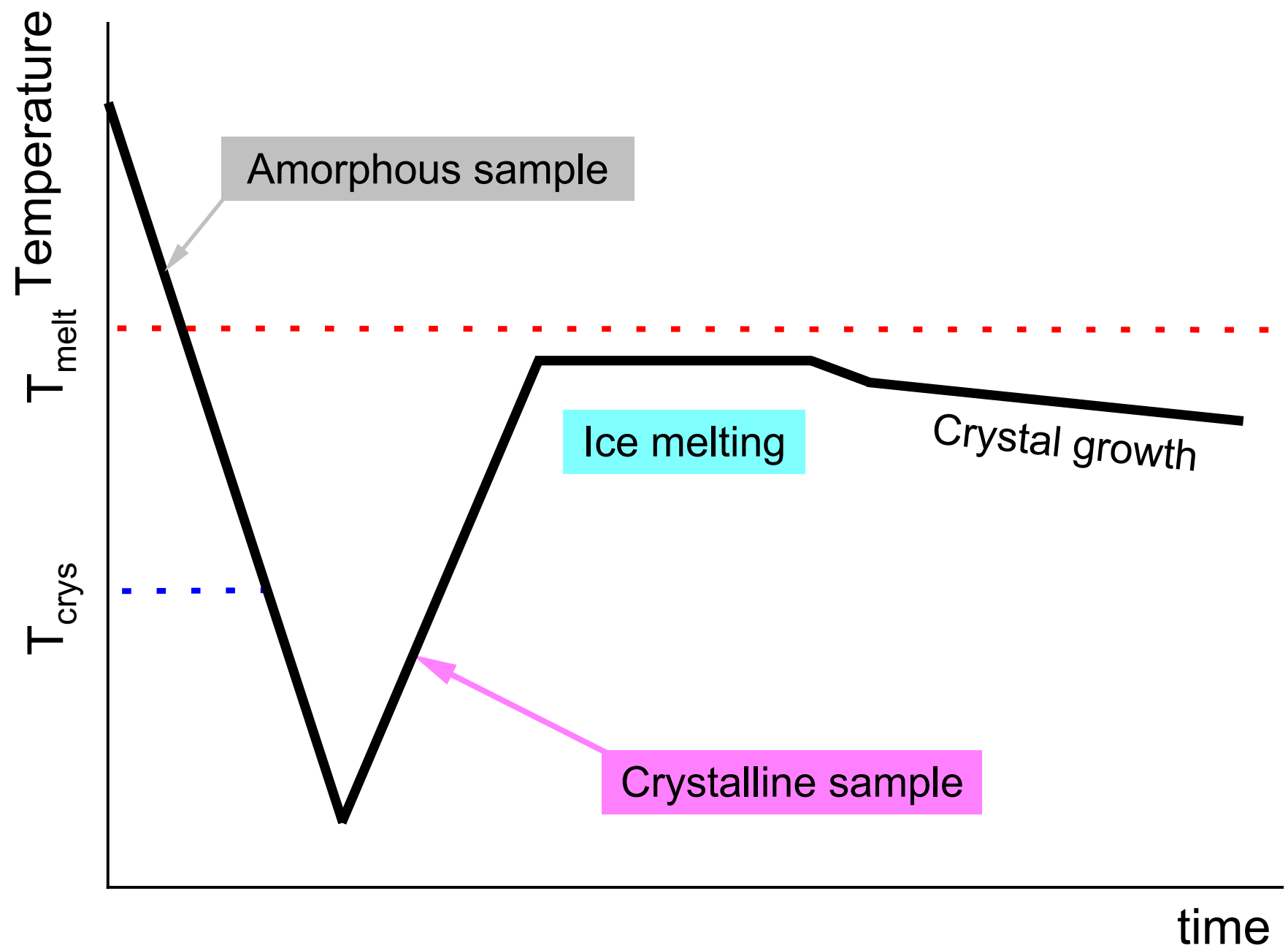

**Figure 1.** Schematic of the temperature protocol used in the kinetic roughening (KR) experiments. Following initial crystallization, the sample was heated into the melting region and held until only a few ice crystals remained. These remaining crystals were then regrown at a controlled rate of 0.1 °C/min, and their morphology was subsequently analyzed. The red dotted line indicates the melting temperature of the solution in the absence of ice. The blue dotted line indicates the temperature of the nucleation of the solution.

The different solutions exhibit varying crystallization and melting temperatures, which we adjusted by altering their concentrations. We cooled the samples rapidly and once the samples crystallized, the temperature was increased to the melting region, and an isotherm was applied until only a few crystals remained; for single-crystal experiments, the isotherm was applied until only one crystal remained. After that, the temperature was gently decreased at a rate of 0.1 ºC/min. Given that the typical diffusion time for water in the DMSO solution is approximately 4 minutes for a 500 μm droplet (assuming a diffusion coefficient of ~0.001 $mm^2/s$)[50], this slow cooling rate allows sufficient time for diffusion to homogenize the solution, minimizing concentration gradients during crystallization. On the other hand, faster cooling rates resulted in rapid growth and consequent dendritic growth. This temperature protocol is schematized in Fig. 1.

For ice recrystallization experiments, after initial freezing, the temperature was elevated and kept constant at the desired temperature.

### 2.3 Data analysis

Real-time brightfield videos were captured using 10X and 50X magnification objectives and a QImaging EXi Aqua digital camera (Teledyne QImaging, Surrey, BC, Canada). The videos were converted into images, from which we analyzed the area, perimeter, and shape of the ice crystals during growth and melting using a MATLAB code (details about the code and the error calculations are provided in the Supplementary Information). Using these values, the change in shape was studied using the roundness function $R$ given in Eq. 1.

$$R = \frac{4\pi A}{P^2} \tag{1}$$

where A and P are the area and perimeter of the crystal, respectively. The roundness function is widely used to quantify the shape of objects in image analysis, with R = 1 for a perfect mathematical circle and R = 0.907 for a perfect regular hexagon. However, when applied to digitally sampled objects, such as those in microscopy images, the measured values are systematically lower due to pixelation, edge discretization, and the specifics of digital perimeter measurement. For example, it was reported a roundness value of 0.911 for a perfect digital circle in ImageJ[51]. This systematic offset should be considered when interpreting digital image analysis results. Similar issues and the necessity to account for measurement errors in digital particle analysis are emphasized in Blott and Pye[52].

## 3. **Results and discussion**

### 3.1 Kinetic roughness transition

As described in the Materials and Methods, the solutions were crystallized and warmed until only a few crystals were left in the field of view. Figure 2 shows representative images of ice crystals grown in DMSO–water solutions under a controlled cooling rate of 0.1 °C/min. This slow cooling rate promotes steady crystal growth without dendritic branching, consistent with near-equilibrium conditions. For $c_{DMSO}$ = 10% (v/v), Fig. 2 (a) and (b), crystals exhibit a round shape from -4.5 to -5 ºC, while for $c_{DMSO}$ = 30% (v/v), Fig. 2 (c) and (d), crystals grow in a hexagonal shape in their basal plane. Importantly, the apparent crystal shape depends on the observed face: circular profiles correspond to basal faces, while rod-like or elongated profiles indicate the prism face. Because the basal face remains faceted up to the melting point and does not undergo roughening, changes in roughness are best monitored by examining crystals oriented with their basal face facing the observer, which reveals transitions in the prism face morphology. Similar results were found for proline solutions in the same temperature range (see Fig. SI1). As anticipated, the roughness transition is determined by temperature rather than the solute.

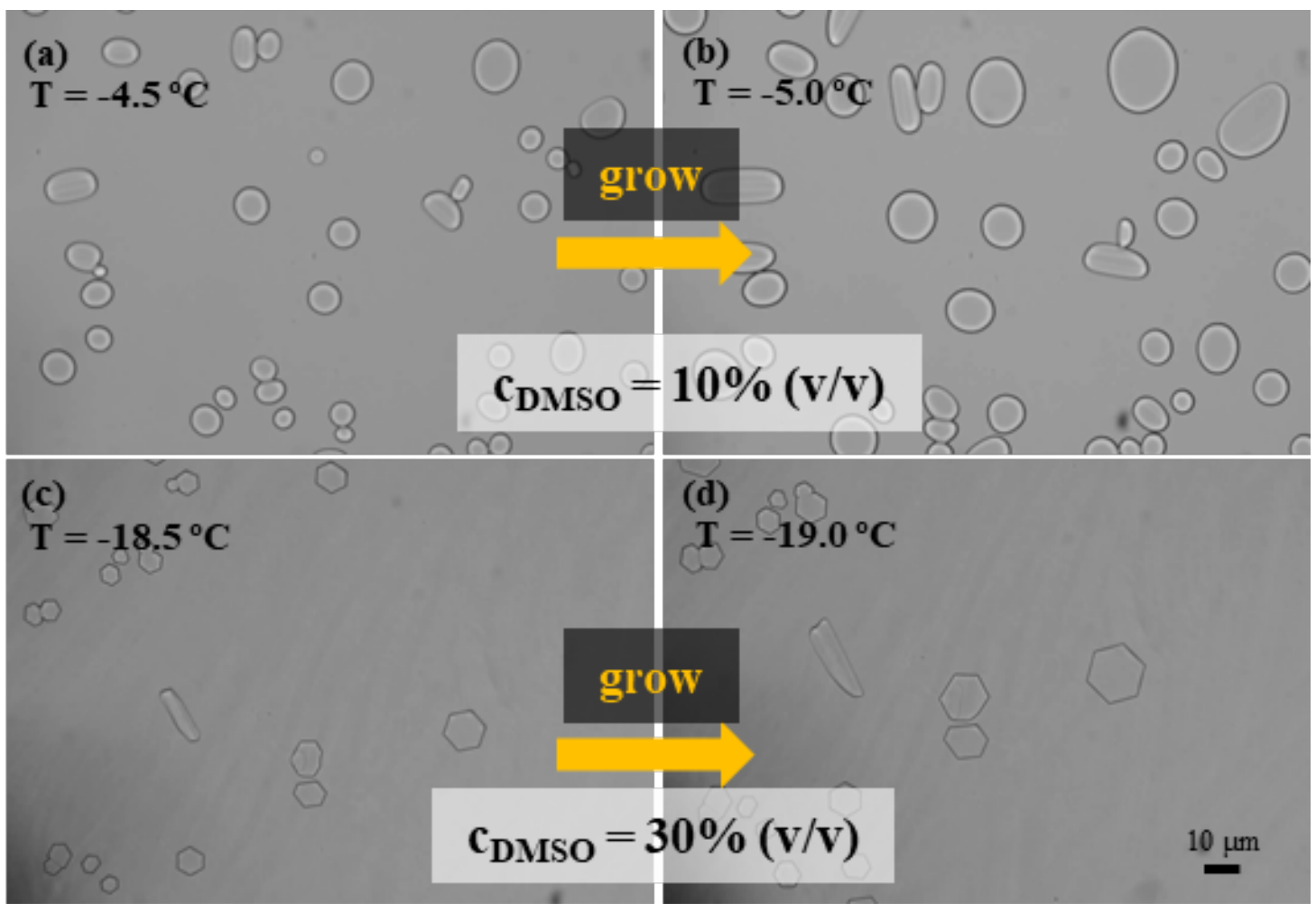

**Figure 2.** Optical microscopy images of ice crystals in DMSO–water solutions at (a, b) 10% and (c, d) 30% DMSO. Samples were slowly cooled at a rate of 0.1 °C/min. For each concentration, two snapshots of the growing crystals are presented, taken five minutes apart. Crystals oriented with their basal face toward the observer (appearing as round or flat hexagons) enable direct monitoring of prism face morphologies. At -5 °C, crystals appear nearly spherical, indicating a rough morphology, while at -19 °C, they develop well-defined hexagonal facets.

To accurately determine $T_R$, we analyzed the shape evolution of individual ice crystals growing in emulsion droplets, where crystal isolation prevents interference from neighboring crystals. This setup also induces near-equilibrium conditions, as solute concentration increases locally during crystal growth, reducing supercooling.

We analyzed three different conditions to determine $T_R$: DMSO emulsions with and without a cover glass (Fig. 3), and in proline solution with a cover glass. Figure 3 shows the evolution of an ice crystal during growth at 0.1 ºC/min in an emulsion of 20% (v/v) DMSO without a cover slip. In the figure, we can distinguish two shapes: a circle corresponding to the droplet solution, and inside it, the ice crystal. The red outline drawn on the contours of the ice crystal corresponds to the shape detection by the MATLAB algorithm. This analysis was performed every minute, i.e., at each 0.1 ºC decrement.

Panels 3 (a-c) provide examples of the crystal morphological evolution at different temperatures during slow cooling: T = -10.5, -15.6, and -18.2 ºC. The ice crystal transitions from a circular disk to a well-defined hexagon as the temperature decreases, indicating a $T_R$ near -15.4 ºC. The crystal also shows physical rotation between Figs. 3 (b) and (c). We speculate that this rotation arises from the growth of facets occurring parallel to the ice surface in the form of advancing steps, which generate torque on the crystal, leading to its rotation. The full time-lapse sequence is provided in Supporting Video SI1.

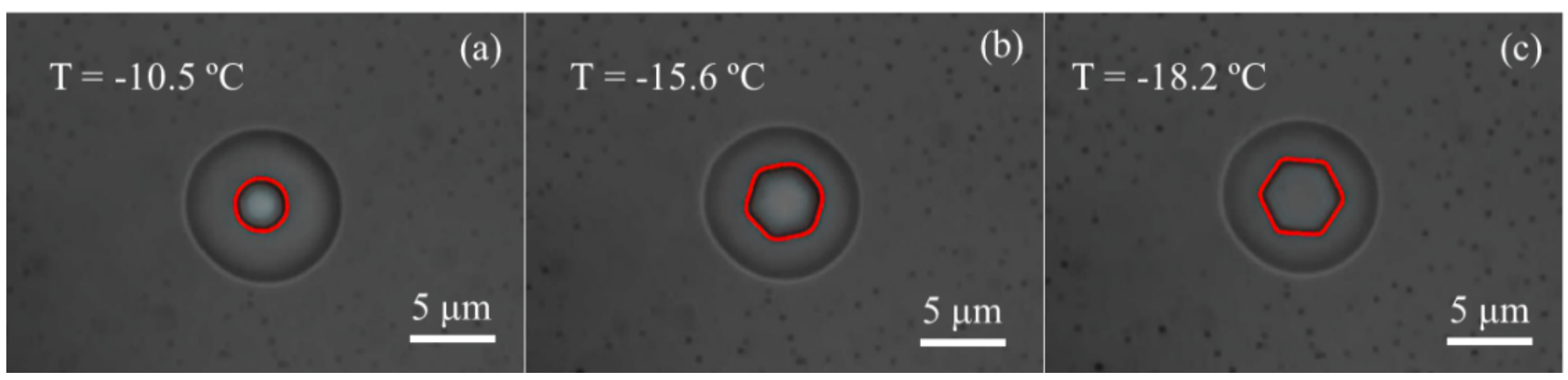

**Figure 3.** Morphological evolution of a single ice crystal in an emulsion of $c_{DMSO}$ = 20% (v/v) during slow cooling. The red outline corresponds to the ice crystal border detected by a MATLAB algorithm. (a) At T = -10.5 ºC, the ice crystal shows a circular shape. (b) At T = -15.6

°C, the crystal grows and changes shape from a disk to an imperfect hexagon. (c) At T = -18.2 °C, the crystal forms a perfect hexagon and shows a rotation relative to its orientation in (b).

As discussed above, the progressive growth of the ice crystal leads to an increase in the solute concentration, since water molecules are continuously incorporated into the solid phase, thereby reducing the volume of liquid water in the system. This increase in solute concentration lowers the melting point of the solution, enabling the system to remain near equilibrium and thus suppressing dendritic growth. In Figure 3 (a-c), the ice crystal occupies approximately 12%, 21%, and 33% of the initial surface area of the confined solution, corresponding to increases in DMSO concentration from 20% (v/v) to approximately 22%, 24%, and 27% (v/v), respectively. Consequently, the melting point decreases from -8.0 °C to -9.4 °C , -10.7 °C, and -13 °C, as calculated following Semenov et al[46]. Further details of these calculations are provided in the section "Estimation of concentration and melting points as a function of crystal size in a drop" in the Supplementary Information. This rise in solute concentration not only depresses the melting temperature but also affects other physicochemical properties of the system, such as the glass transition temperature[53].

Figure 4 shows the quantitative roundness analysis of the ice crystal featured in Supporting Video SI1, plotted as a function of temperature. For each video frame, the crystal boundary was extracted, and roundness was calculated using Equation 1. The resulting curve was fitted with a sigmoidal function (Eq. 2).

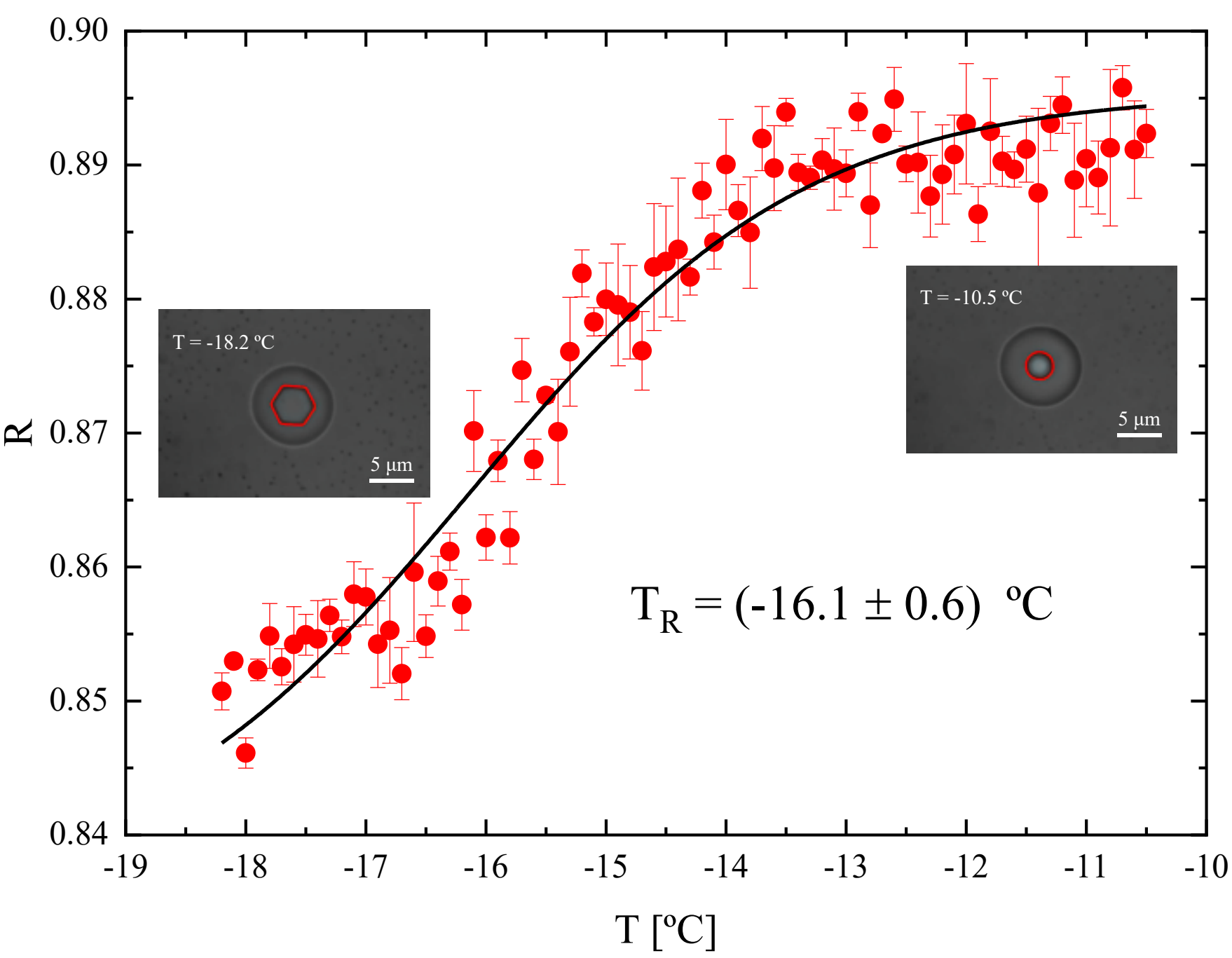

**Figure 4.** Roundness of a single ice crystal as a function of temperature during slow cooling in $c_{DMSO}$ = 20% (v/v) emulsion. The crystal was grown at a cooling rate of 0.1 °C/min in oil and imaged at 50× magnification. Each data point shows the crystal's shape from one video frame. The black line is a sigmoidal fit, with $T_R$ indicated by the curve's inflection point.

$$R_{grow}(T) = R_0 - \frac{R_{1,grow}}{1+e^{\frac{(T-T_R)}{\sigma}}} \tag{2}$$

where $R_0$ is the roundness of a disk, $R_{1,grow}$ is the amplitude of the sigmoidal, σ is the stretching factor and $T_R$ corresponds to the roughness transition temperature. For $c_{DMSO}$ = 20% (v/v) without a cover slip, $T_R$ was (-16.1 ± 0.6) °C, while with a cover glass, $T_R$ was (-15.7 ± 0.2) °C (Fig. SI2 (a)). For 30% (w/w) proline solution emulsion with a cover glass, $T_R$ was (-16.2 ± 0.2) °C (Fig. SI2 (b)). Taking these measurements together, we report a consistent $T_R$ of (-16.0 ± 0.2) °C. This value aligns well with previous studies of the KR under high pressure conditions[10–12]. Table 2 reports the averages of the Eq. 2 fit; based on these, the model yields a 10-90% transition of -18.8 to -13.2 °C.

**Table 2.** Average parameters obtained from fitting Eq. (2) to the growth roundness data. $R_0$ is the roundness of a disk, $R_{1,grow}$ is the amplitude of the sigmoidal, σ is the stretching factor, and $T_R$ is the KR temperature.

| Parameters | Value |
|---|---|
| $R_0$ | 0.90 ± 0.01 |
| $R_{1,grow}$ | 0.062 ± 0.001 |
| σ | 1.26 ± 0.06 ºC |
| $T_R$ | -16.0 ± 0.2 °C |

This observation supports the idea that the KR is governed primarily by temperature, not by the specific solute or applied pressure[5]. In our system, proline and DMSO reduce the melting point but do not interact with the ice interface. Their role is to suppress rapid growth by lowering the degree of supercooling, not to affect crystal morphology through adsorption. It should be noted that in any solution where the solute does not interact directly with the ice interface, such as DMSO, proline, glycerol, or fructose, the KR is expected to be governed solely by temperature. Once the melting point is depressed to temperatures below $T_R$, crystals at slightly supercooled temperatures will grow faceted

As mentioned in the introduction, some materials exhibit IRI effects without significant TH but with noticeable ice-shaping properties. For example, cellulose nanocrystals in sucrose solutions have demonstrated strong IRI and ice-shaping effects, with the shape of ice crystals transitioning from disks to hexagons at -8 ºC after 2 hours under isothermal conditions[54,55]. Similarly, zirconium acetate solutions show varying ice crystal morphologies, such as elongated hexagonal tubes, depending on the solution's pH, while buffer solutions yield disk-shaped crystals[16]. These findings suggest that these solutes interact with ice crystals to some extent. A similar conclusion was drawn by Dreischmeier *et al.*, who studied the ice-shaping capacity of boreal pollen[55].These experiments were done at temperatures well above the $T_R$. Therefore, the presence of faceting at lower temperatures, i.e., below $T_R$, cannot be used as conclusive evidence of ice-binding activity. Our findings emphasize that intrinsic KR can produce faceted morphologies in the complete absence of molecular adsorption.

Until this point, our analysis has focused on ice crystal growth. We now examine the melting behaviour and its differences from growth. Figure SI3 (a-c) shows the morphological evolution of a single crystal during melting. Beginning as a well-defined hexagon at -19 ºC, the crystal progressively transforms into a circular disk. The rotation observed between Figs. SI3(a) and (b) was previously explained in detail by Pertaya *et al*[37] and Cahoon *et al*[12]. This rotation, contrary to the physical rotation of the crystal observed during growth and potentially related to the step growth of the facets, is not an actual rotation but rather a consequence of the preferential melting of the corners.

While the crystal's roundness appears to change similarly to what was observed during growth (see Figure 4), the underlying kinetics differ. Growth follows a sigmoidal roundness–temperature dependence, indicative of KR. In contrast, melting proceeds with an exponential relaxation, described by Equations 3 (a) and (b).

$$R_{melt}(T) = R_0 - R_{1,melt}\, e^{-T/\sigma_m} \tag{3(a)}$$

$$R_{melt}(t) = R_0 - R_{1,melt}\, e^{-t/\tau} \tag{3(b)}$$

where $R_0$ is the roundness of a disk, $R_{1,melt}$ is the amplitude of the exponential, and $\sigma_m$ and $\tau$ are the characteristic decay constants for gradual melting (as a function of temperature) and isothermal melting after sudden temperature increase (as a function of time), respectively. This behaviour is consistent during melting, regardless of the temperature, unlike the roughness transition, where sigmoidal behaviour is observed only between -10 and -20 ºC. Crystals were melted in two distinct ways: (1) by gradually warming, analogous to the slow cooling used for growth, yielding $\sigma_m$ = (3.5 ± 0.4) ºC (Fig. 5a); and (2) by holding the temperature constant and observing shape evolution over time, yielding τ = (47.3 ± 3.7) s (Fig. 5b). In both cases, the crystal evolved from a hexagonal to a circular morphology. Notably, while time is the independent variable in the isothermal melting experiments, in the gradual warming case, time correlates with temperature, making both representations valid under Equations 3. These results highlight a key finding: growth and melting proceed through fundamentally different morphological pathways, reinforcing the asymmetry in interface kinetics introduced earlier. As noted in the introduction, shaping during melting is possible in ice-active materials, as demonstrated for hyperactive antifreeze proteins that modulate ice crystal morphology during melting[16,38]

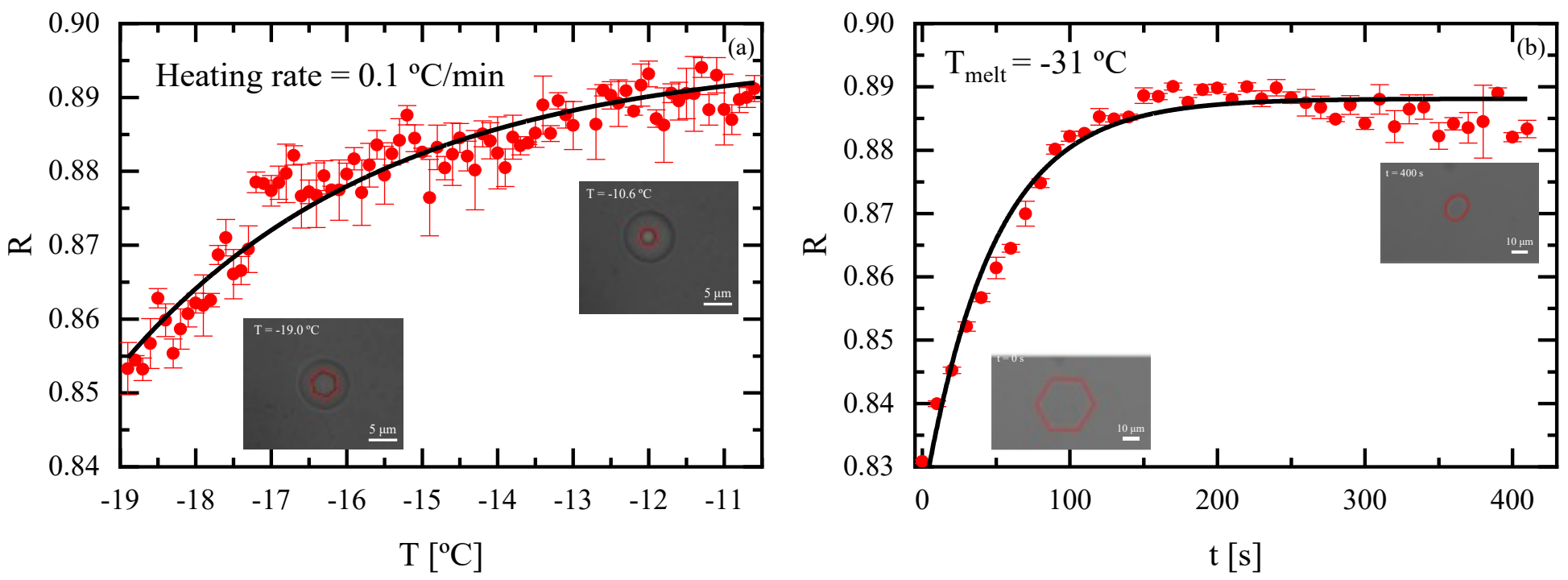

**Figure 5.** Roundness evolution of a single ice crystal during melting, shown for two conditions: (a) for emulsions of $c_{DMSO}$ = 20% (v/v) and gradual warming at 0.1 °C/min, and (b) for emulsions of $c_{DMSO}$ = 40% (v/v) and isothermal melting at constant temperature of -31 ºC. In both cases, the analysis tracks the shape change as a hexagonal crystal becomes circular.

3.2 Roughness transition during ice recrystallization

Video SI2 shows the ice recrystallization of a DMSO solution with $c_{DMSO}$ = 20% (v/v) at -9 °C over 5 hours. It is well known that under isothermal conditions applied to a crystallized aqueous solution, larger crystals grow at the expense of smaller ones, as observed in the video. During the ice recrystallization above $T_R$, the crystals maintain a round shape without developing distinct right angles. A similar behavior is observed in the presence of AFPIII at 0.1 µM in $c_{DMSO}$ = 20% (v/v).

However, when the AFPIII concentration increases to 0.5 µM, a notable difference emerges. In Video SI3 it is possible to distinguish how ice crystals tend to adopt a hexagonal shape during growth, whereas they assume a round shape during the melting process. AFPIII altered crystal morphology through a distinct mechanism that overrides the effects of temperature alone. Notably, the ability of AFPIII to maintain faceted morphologies at temperatures above the roughening transition indicates that antifreeze proteins suppress KR. At higher concentrations, such as 1 μM, thermal hysteresis remains nearly negligible, but IRI persists, preventing for further analysis. A detailed explanation of this phenomenon is provided below.

Figure SI4 shows a crystallized solution of $c_{DMSO}$ = 30% (v/v) held isothermally at -20 °C for 1.5 hours. During this period, some crystals grew while others melted, allowing morphological changes to be assessed. Above $T_R$, both disk-like and hexagonal crystals coexisted, an effect also seen in Video SI4 and mirrored in Video SI3 for 0.5 μM AFPIII in $c_{DMSO}$ = 10 % (v/v) at -9 ºC. However, this should not be interpreted as a genuine roughness transition, as the presence of AFPIII interferes with ice crystal morphology by modifying the probability of step nucleation. This pinning effect stabilizes facets at higher temperatures, distinguishing ice shaping by AFPs from a temperature dependent roughness transition. These findings highlight that ice shaping induced by ice-active materials, such as AFPs or cellulose nanocrystals, is mechanistically distinct from intrinsic KR.

After 37 minutes at -20 ºC, isolated single crystals were visible within the sample, permitting more detailed shape analysis. Figure SI4 shows the state of the sample at the beginning of the isotherm and after 58 minutes of the isothermal hold. These images were taken with a 10X objective, while the images shown in Figure 6 were captured with a 50X objective. The red square in Figure SI4 corresponds to the section analyzed in Figure 6. One of the major differences between single crystal experiments (see the previous section) and ice recrystallization experiments is that, in the former, crystal growth was initiated from a disk, and the crystal growth was controlled and monitored. In contrast, during recrystallization, crystals often exhibit irregular shapes and may fuse over time. This fusion leads to neck formation between adjacent crystals, a feature observed in previous studies of ice recrystallization[56,57]. Future work could investigate whether the presence of ice-binding proteins modifies neck formation dynamics, potentially altering crystal coalescence and stability.

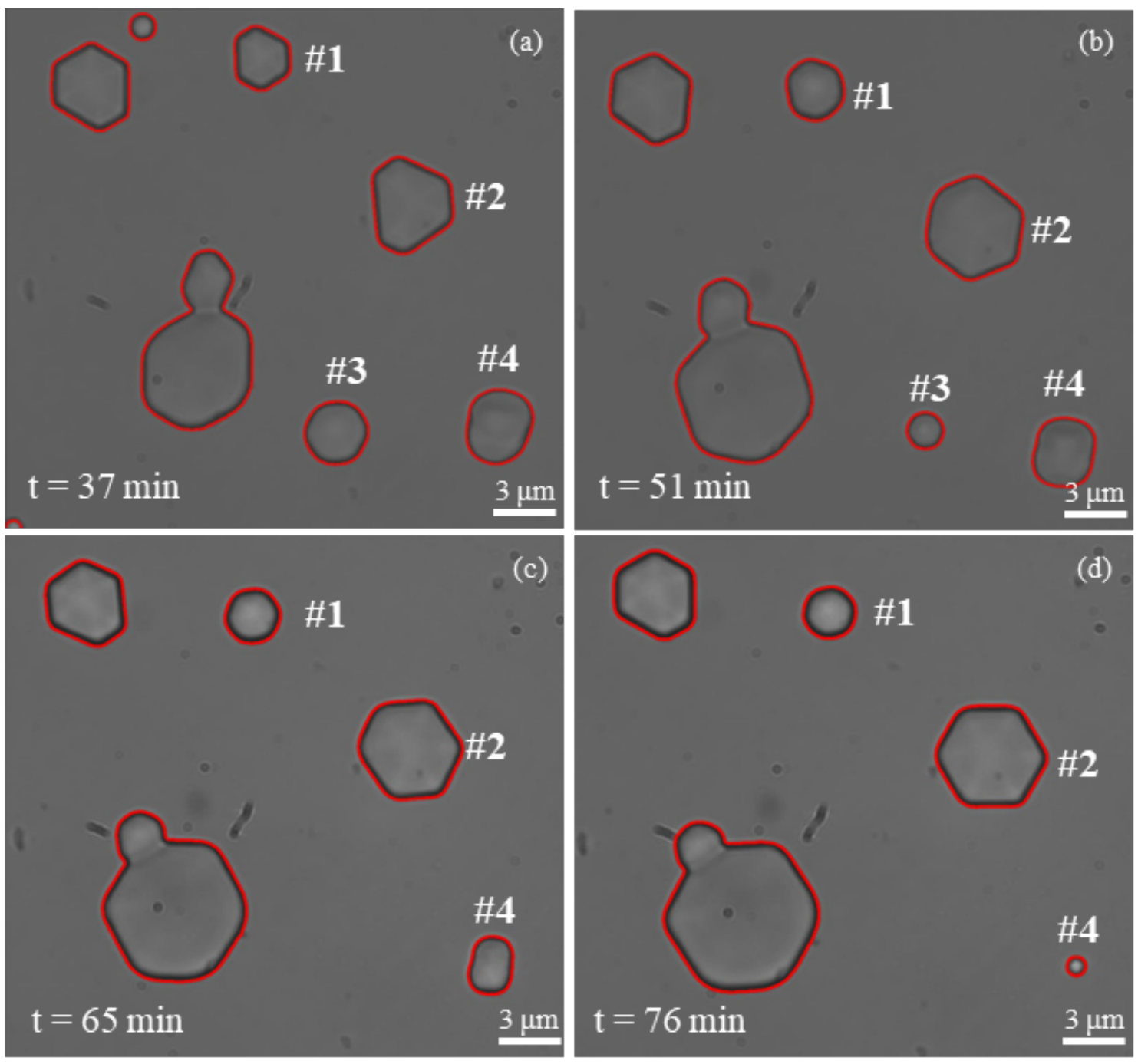

**Figure 6.** Recrystallization of ice at -20 °C for 1.5 hours in a solution with 30% (v/v) DMSO. As crystals grow, they acquire well-defined hexagonal shapes; as they melt, they lose facets and become increasingly circular. Red outlines identify and track individual crystals, revealing the asymmetric evolution of roughness between growth and melting. The field of view is the red-marked region from Figure SI4, shown here at 50× magnification.

Figure 6 shows the morphological evolution of ice crystals during recrystallization, distinguishing between those that grow and those that melt. Although it may not be possible to perform an analysis as detailed as in the single crystal case, some observations can be made. Crystals that melt tend to exhibit a circular shape. For instance, crystal #1 initially presents an irregular hexagonal geometry but gradually transforms into a rounded disk as it melts. In contrast, crystals that are growing tend to become increasingly faceted. The crystal #2, which begins with a distorted hexagonal form, evolves into a well-defined hexagon. Meanwhile, the crystals #3 and #4 show progressive rounding during melting.

This tendency toward hexagonal symmetry during growth likely reflects a preferential advancement of existing facets, driven by enhanced step nucleation and propagation. In contrast, the loss of faceting during melting, particularly at corners, is consistent with interface smoothing as step edges retract.

The crystal #1 is particularly interesting to follow, as it begins with an irregular hexagonal shape and gradually becomes circular before disappearing. Figure 7 shows the roundness function analysis of this crystal, demonstrating how the roundness function approaches a value of 0.89, indicating a circular shape.

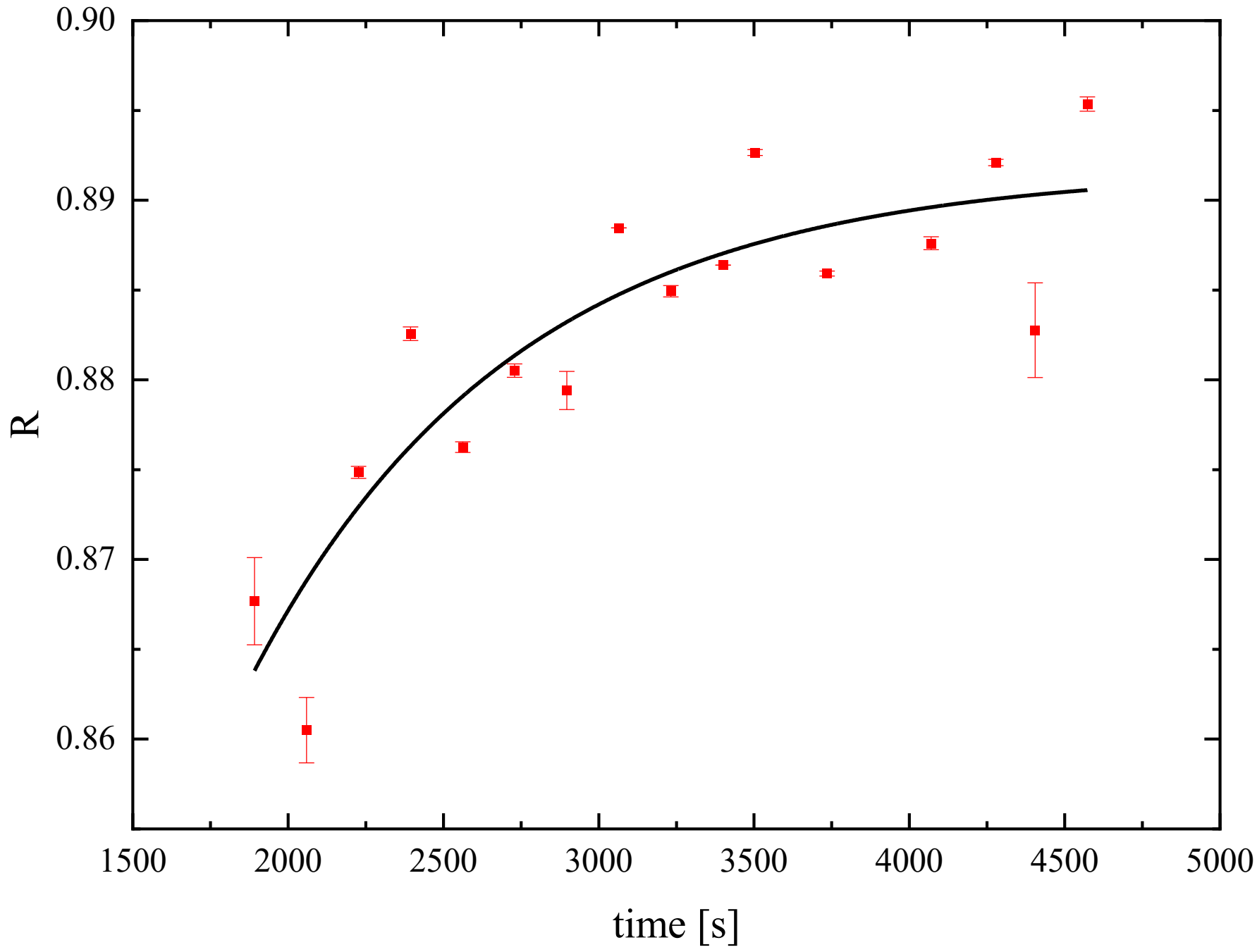

**Figure 7.** Roundness evolution of the crystal #1 from Figure 6 during melting. The crystal transitions from an irregular hexagonal shape to a nearly circular form, consistent with an exponential decay described by Equations 3. The black line shows the Eq. 3 fit, yielding $\tau$ = (846 ± 210) s. This behaviour reflects the loss of faceting during melting, in contrast to the sigmoidal trend observed during growth.

Williamson et al. studied ice rippling in fructose solutions at -17 to -20 ºC[6]. They observed hexagonal crystal shapes during growth and suggested that the equilibrium Wulff shape had been reached. However, we propose that under recrystallization conditions, there remains a driving force for growth; the system, while close to equilibrium, is not fully equilibrated, and the observed faceting results from kinetic roughening.

In a recent study, IRI experiments conducted at low temperatures briefly mention the observation of hexagonal crystals, but offer no detailed analysis or explanation[47]. Moreover, most IRI studies are conducted at temperatures between 0 and -8 °C, above $T_R$. In that regime, any changes in crystal morphology are likely due to solute-ice interactions rather than intrinsic KR. By contrast, when recrystallization experiments are conducted below -16 °C, the distinct transition in crystal shape, i.e., from round during melting to hexagonal during growth, becomes clearly observable. This work provides a comprehensive analysis of this behaviour, emphasizing the importance of temperature-dependent KR in interpreting ice morphology during recrystallization.

## 4. Conclusion

This study establishes that the KR of ice crystals, previously reported under high-pressure conditions[10,11,58], occurs at a nearly identical temperature (-16.0 ± 0.2 °C) in proline-water and DMSO solutions at atmospheric pressure. It is consistent with the finding that ice in a high concentration of fructose grows in faceted hexagonal[6] at temperature lower than the $T_R$ we found. This confirms that KR is primarily temperature-governed, independent of both solute identity and pressure. Importantly, these results generalize the universality of KR behavior in ice-liquid systems. Our findings advance prior understanding in several critical aspects:

- New conceptual distinction: We demonstrate that ice morphology can exhibit faceting via intrinsic KR, even in the absence of ice-binding molecules. This decouples crystal morphology from the presence of ice-active solutes. Below ≈ -12 °C, we recommend explicitly checking for KR contributions when interpreting faceted ice morphologies, since intrinsic faceting can arise even without ice-active molecules. Based on the Eq. 2 fit parameters (Table 2), the 10-90% transition spans -18.8 to -13.2 °C.
- Asymmetric growth and melting kinetics: Growth kinetics followed the sigmoidal form of Eq. 2, with $\sigma \approx 1.26$ °C transition from no barrier of new layers to layer-propagation-limited advance under near-equilibrium conditions. In contrast, melting obeyed the exponential dependence of Eqs. 3, independent of temperature, and consistently drove crystals toward rounded morphologies. The decay constant varied strongly with the melting pathway: $\sigma_m \approx 3.5$ °C for gradual warming (2100 s process time), $\tau \approx 47.3$ s for isothermal melting after sudden temperature increase, and $\tau \approx 846$ s for recrystallization-driven melting via Ostwald ripening. The persistence of rounding during melting, in sharp contrast to the hexagonal faceting observed during growth below $T_R$, reveals a pronounced asymmetry in interface kinetics. This previously unreported asymmetry between growth- and melting-induced morphological development advances interface science by establishing that nonequilibrium melting dynamics follow distinct kinetic regimes from those governing growth.
- Interaction with ice-active molecules: AFPIII at 0.5 µM, a concentration corresponding to ~50% IRI[59], modulates faceting above $T_R$ and elevates the roughening transition temperature, which we speculate occurs via step stabilization. In contrast, at 0.1 µM AFPIII no ice shaping was observed. Together, these findings suggest that the thresholds for ice shaping and recrystallization inhibition coincide, underscoring their shared sensitivity as indicators of ice activity.

The broader implications of this work span cryobiology, where KR may impact optimization of vitrification protocols; atmospheric science, where ice nucleation and morphology affect cloud radiative properties; and interface engineering, where molecular control of surface roughness transitions may be exploited to design advanced antifreeze coatings, ice-resistant materials, or biomimetic crystallization modulators.

Future directions include:

- Investigate the prism-face symmetrization of hexagonal crystals during recrystallization below $T_R$, where initially anisotropic growth rates gradually converge until all facets advance in unison.
- Investigate crystal rotations during growth across the roughening transition using different salts and complementary simulations.
- Molecular-level investigation of roughening suppression by different classes of ice-active agents.
- Extension of KR studies to other inorganic and organic crystallizing systems exhibiting faceting transitions at solid-liquid interfaces.

**Acknowledgment**

We are thankful for the support from the Israel Science Foundation and NOFAR program of the Israel Innovation Authority.

**Figure captions**

**Figure 2.** Schematic of the temperature protocol used in the kinetic roughening (KR) experiments. Following initial crystallization, the sample was heated into the melting region and held until only a few ice crystals remained. These remaining crystals were then regrown at a controlled rate of 0.1 °C/min, and their morphology was subsequently analyzed. The red dotted line indicates the melting temperature of the solution in the absence of ice. The blue dotted line indicates the temperature of the nucleation of the solution.

**Figure 2.** Optical microscopy images of ice crystals in DMSO–water solutions at (a, b) 10% and (c, d) 30% DMSO. Samples were slowly cooled at a rate of 0.1 °C/min. For each concentration, two snapshots of the growing crystals are presented, taken five minutes apart. Crystals oriented with their basal face toward the observer (appearing as round or flat hexagons) enable direct monitoring of prism face morphologies. At -5 °C, crystals appear nearly spherical, indicating a rough morphology, while at -19 °C, they develop well-defined hexagonal facets.

**Figure 3.** Morphological evolution of a single ice crystal in an emulsion of $c_{DMSO}$ = 20% (v/v) during slow cooling. The red outline corresponds to the ice crystal border detected by a MATLAB algorithm. (a) At T = -10.5 ºC, the ice crystal shows a circular shape. (b) At T = -15.6 ºC, the crystal grows and changes shape from a disk to an imperfect hexagon. (c) Finally, at T = -18.2 ºC, the crystal forms a perfect hexagon and shows a rotation relative to its orientation in (b).

**Figure 4.** Roundness of a single ice crystal as a function of temperature during slow cooling in $c_{DMSO}$ = 20% (v/v) emulsion. The crystal was grown at a cooling rate of 0.1 °C/min in oil and imaged at 50× magnification. Each data point shows the crystal's shape from one video frame. The black line is a sigmoidal fit, with $T_R$ indicated by the curve's inflection point.

**Figure 5.** Roundness evolution of a single ice crystal during melting, shown for two conditions: (a) for emulsions of $c_{DMSO}$ = 20% (v/v) and gradual warming at 0.1 °C/min, and (b) for emulsions of $c_{DMSO}$ = 40% (v/v) and isothermal melting at constant temperature of -31ºC. In both cases, the analysis tracks the shape change as a hexagonal crystal becomes circular.

**Figure 6.** Recrystallization of ice at -20 °C for 1.5 hours in a solution with 30% (v/v) DMSO. As crystals grow, they acquire well-defined hexagonal shapes; as they melt, they lose facets and become increasingly circular. Red outlines identify and track individual crystals, revealing the asymmetric evolution of roughness between growth and melting. The field of view is the red-marked region from Figure SI4, shown here at 50× magnification.

**Figure 7.** Roundness evolution of the crystal #1 from Figure 6 during melting. The crystal transitions from an irregular hexagonal shape to a nearly circular form, consistent with an exponential decay described by Equations 3. The black line shows the Eq. 3 fit, yielding $\tau$ = (846 ± 210) s. This behaviour reflects the loss of faceting during melting, in contrast to the sigmoidal trend observed during growth.

## Tables

**Table 1.** Melting temperatures of DMSO and proline aqueous solutions at various concentrations.

| | DMSO [% (v/v)] | | | | | Proline [% wt] | | |
|---|---|---|---|---|---|---|---|---|
| | 10 | 20 | 25 | 30 | 40 | 20 | 30 | 40 |
| Melting Temperature [ºC] | -3.5 (-3.2) | -8.7 (-8.0) | -10.6 (-11.3) | -17.7 (-15.6) | -31.0 (-29.0) | -6.5 | -10.5 | -16.4 |

**Table 2.** Average parameters obtained from fitting Eq. (2) to the growth roundness data. $R_0$ is the roundness of a disk, $R_{1,grow}$ is the amplitude of the sigmoidal, σ is the stretching factor, and $T_R$ is the KR temperature.

| Parameters | Value |
|---|---|
| $R_0$ | 0.90 ± 0.1 |
| $R_{1,grow}$ | 0.062 ± 0.001 |
| σ | 1.26 ± 0.06 ºC |
| $T_R$ | -16.0 ± 0.2 °C |

**Supplementary Information of**
**Roughness transition and its implication during ice recrystallization**

Jorge H. Melillo[1] and Ido Braslavsky[1]

[1]Institute of Biochemistry, Food Science, and Nutrition, Robert H. Smith Faculty of Agriculture, Food and Environment, The Hebrew University of Jerusalem, Rehovot 7610001, Israel

In this supplementary we included the code for image analysis of the videos, and figures of the proline results and additional results for the DMSO solutions. Videos SI1–SI4 provide dynamic visualizations supporting these findings, including facet development, melting behavior, and recrystallization dynamics.

## MATLAB code for crystal morphological detection

```
Folder = '%(folder where the images are located)';
Files = dir([Folder '*.jpg']);
out_folder = '%(folder where the images will be saved)';

areas = zeros(1, numel(Files));
perimeters = zeros(1, numel(Files));

for i = 1:numel(Files)
      Image = imread(fullfile(Folder, Files(i).name));
      Image_grey = rgb2gray(Image); %change color to grey picture
      Filter = adaptthresh(Image_grey, 0.85); % contrast filter
      Filtered_image = imbinarize(Image_grey, Filter);

      Edge = edge(Filtered_image, 'Canny'); % border detection
      Contour = bwboundaries(Edge); % Find shapes

      max_area = 0;
      max_perimeter = 0;
      max_contour = [];

      for k = 1:length(Contour)
            Boundary = Contour {k};
            area = polyarea(Boundary(:,2), Boundary(:,1)); % area
      calculation
            perimeter = sum(sqrt(diff(Boundary(:,2)).^2 +
            diff(Boundary(:,1)).^2)); % perimeter calculation

            if area > max_area
                  max_area = area;
                  max_perimeter = perimeter;
                  max_contour = Boundary;
            end
      end

      areas(i) = max_area;
      perimeters(i) = max_perimeter;

      % Shape drawing
      imshow(Image); hold on;
      plot(max_contour (:,2), max_contour (:,1), 'r', 'LineWidth', 2);
      hold off;

      Output = fullfile(out_folder, Files(i).name);
      saveas(gcf, Output);
```

```
end

T = -10.5:-0.1:-18.2;
R = 4*pi*areas./perimeters.^2;
Data = [T' , R', areas', perimeters'];
dlmwrite('Data.txt',Data,'delimiter', ' ');
```

**Influence of Pixelation and Thresholding on Roundness Measurement**

In digital image analysis, two primary sources affect the quantification of roundness: pixelation and thresholding sensitivity. Pixelation is an inherent property of the imaging system, determined by the pixel size and grid, which causes even a perfect circle to be represented as a "staircased" object with discrete steps rather than a smooth curve. As a result, the maximum roundness value attainable for a perfect circle is systematically reduced from the theoretical value of 1 to approximately 0.90–0.91[1] in typical high-resolution digital images, as explained in 2.3 of the article. Thus, pixelation defines the baseline offset for roundness in digital measurements.

In addition to pixelation, thresholding during image segmentation introduces further variability. The threshold parameter determines which pixels are classified as part of the object or the background, and small changes can shift the detected boundary by one or more pixels around the perimeter. This causes fluctuations in the calculated area, perimeter, and roundness, resulting in a range of roundness values for the same object, typically centered around the pixelation-defined baseline. For example, in our analysis, a perfect digital circle might exhibit a roundness of 0.90 due to pixelation, but the measured value may fluctuate between 0.88 and 0.91 depending on the chosen threshold. Unlike pixelation, which is constant for a given system, thresholding error can vary between images and depends on image contrast and edge definition.

To estimate the uncertainty in roundness due to thresholding, we analyzed each image using three threshold values (0.83, 0.85, and 0.87). In our code, the line that corresponds to the threshold is **`Filter = adaptthresh(Image_grey, 0.85)`**. The final roundness value was taken as the mean of the results from these three thresholds, and the associated uncertainty was estimated as the standard deviation across these measurements.

**Estimation of concentration and melting points as a function of crystal size in a drop**

As described in the main manuscript, as ice crystals grow within a confined solution, the solute concentration in the remaining liquid phase increases due to the exclusion of solute from the growing ice. The MATLAB algorithm developed here allows us to determine the area of the ice crystal during growth, and, using similar code, we can easily measure the total area of the confined solution before the ice nucleation. This enables calculation of the fraction of frozen water as:

$$f_{frozen} = \frac{A_{ice}}{A_{total}}$$

where $A_{ice}$ is the area of the ice crystal and $A_{total}$ is the area of the confined solution (before the ice nucleation). As the preperation is flat, the area ratio represents also the volume fraction. Assuming that only water molecules are incorporated into the growing crystal and that solutes are excluded, the local solute concentration increases as freezing progresses. The final concentration of liquid water in the concentrated solution can be estimated as:

$$V_{H2O,liq,fin} = V_{H2O,liq,in} - V_{ice} = V_{H2O,liq,in} - V_{H2O,liq,in} * f_{frozen} = V_{H2O,liq,in}\,(1 - f_{frozen})$$

where $V_{H2O,liq,in}$ is the initial water volume in the solution (before ice nucleation and growth), and $V_{ice}$ is the volume of water incorporated into the ice phase. Since the amount of DMSO remains constant throughout ice growth, its new concentration in the liquid phase can be calculated as:

$$c_{DMSO,fin} = \frac{V_{DMSO}}{V_{DMSO} + V_{H2O,liq,fin}}$$

With this updated concentration, it is possible to estimate the melting point using the phenomenological fit obtained by Semenov et al[2]. It is important to note that our concentrations are in volume/volume (v/v), while Semenov et al. report concentrations in weight percent (wt%). Therefore, a conversion using the densities of water and DMSO is required prior to these calculations.

Finally, we note that these are estimations based on several simplifying assumptions: (i) the volumes of ice and liquid water are considered equal, (ii) the area of the emulsion droplet does not change during the experiment, and (iii) there is no significant engulfment of DMSO in the ice phase. Despite these approximations, this method provides a reasonable estimation of how the growth of the ice crystal increases the DMSO concentration in the remaining liquid solution.

**Figures**

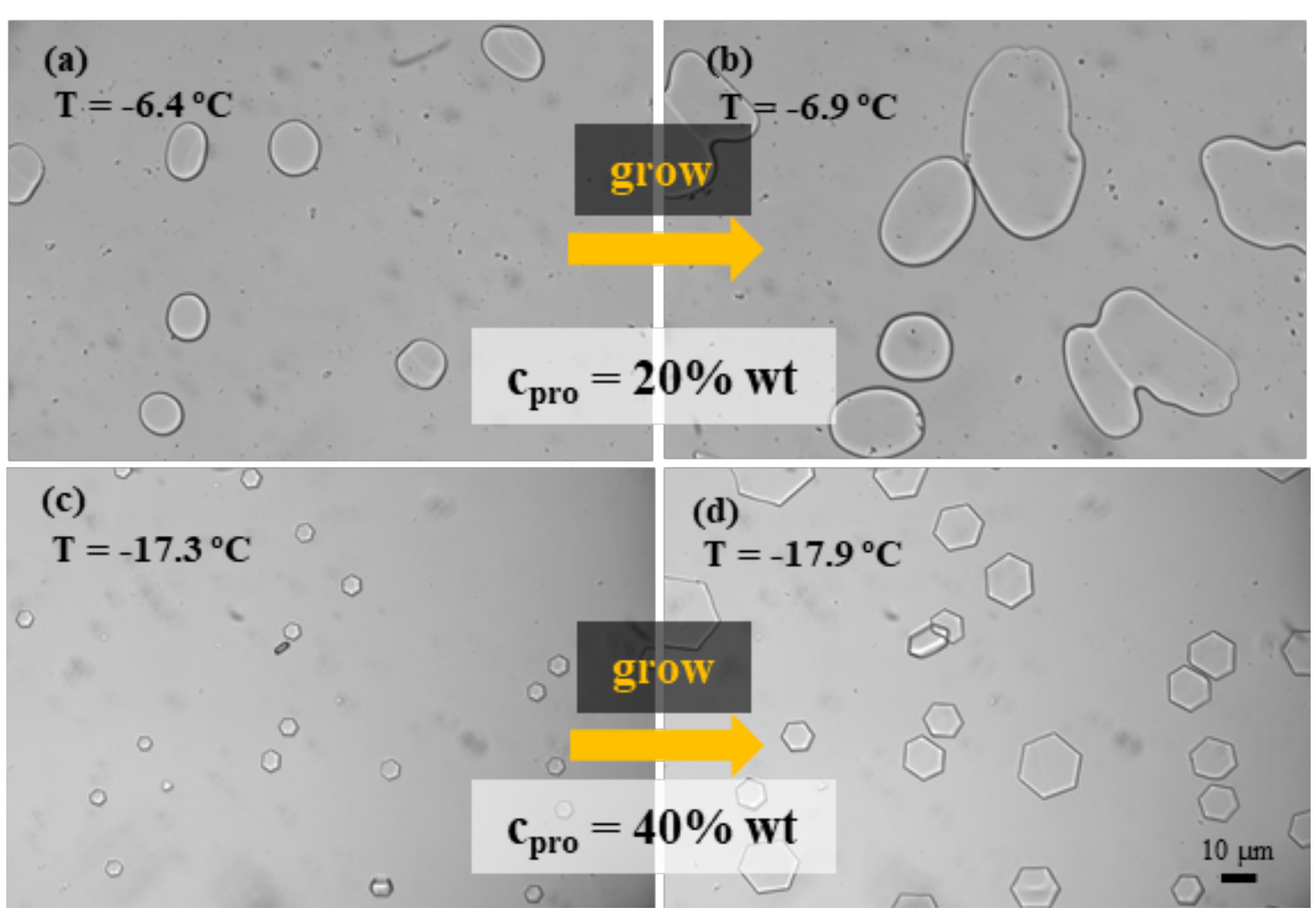

**Figure SI1.** Optical microscopy images of ice crystals in proline–water solutions at (a, b) 20% and (c, d) 40% proline. Samples were slowly cooled at a rate of 0.1 °C/min. For each concentration, two snapshots of the growing crystals are shown. Crystals oriented with their basal face toward the observer (appearing as round or flat hexagons) enable direct monitoring of prism face morphologies. At -6.9 °C, crystals appear nearly spherical, indicating a rough morphology, while at -17.9 °C, they develop well-defined hexagonal facets.

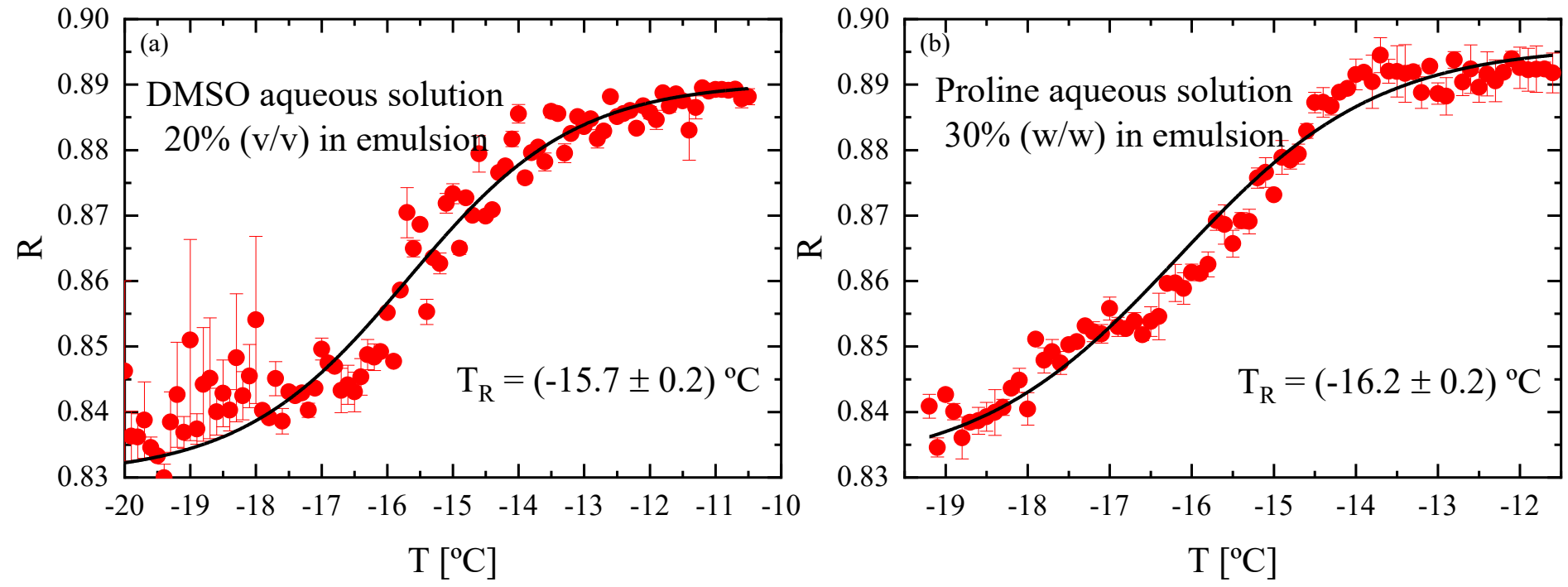

**Figure SI2.** Roundness as a function of temperature for individual ice crystals growing in emulsion droplets. The black line represents a sigmoidal fit, from which $T_R$ is extracted as the inflection point. (a) Emulsion of 30% (w/w) proline solution enclosed between a glass cover. (b) Emulsion of 20% (v/v) DMSO solution without a glass cover. These measurements reveal a consistent TR near -15.4 °C, independent of solute type and boundary conditions.

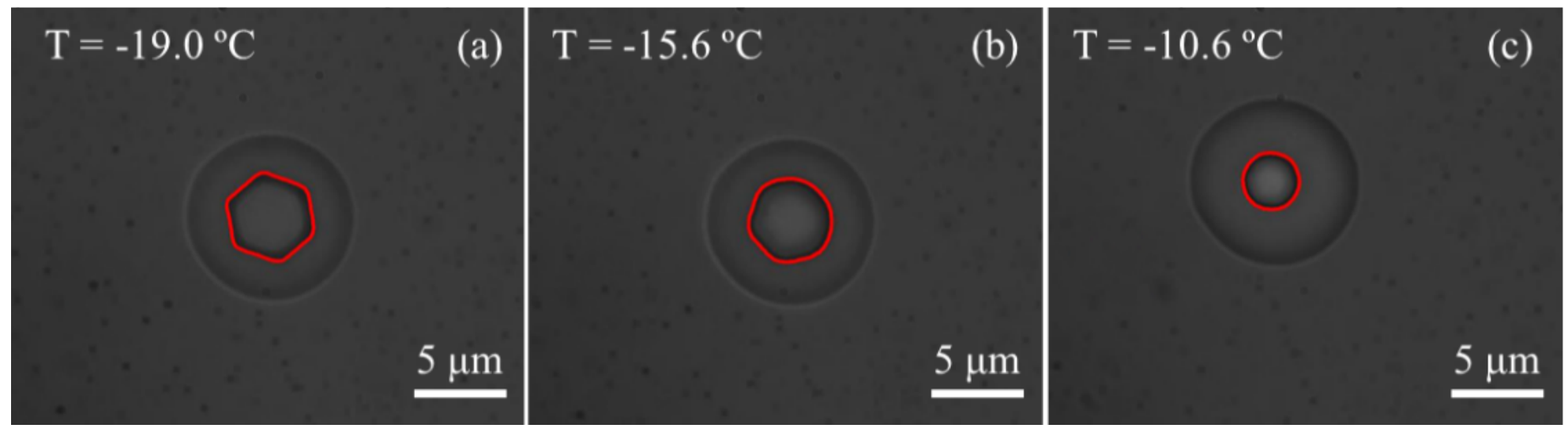

**Figure SI3.** Melting evolution of a single ice crystal. (a) The crystal begins melting from an initially well-defined hexagonal shape. (b) Faceted edges progressively disappear, and an apparent rotation is observed due to preferential melting at the corners. (c) The crystal evolves into a rounded disk-like shape, consistent with surface roughening during melting.

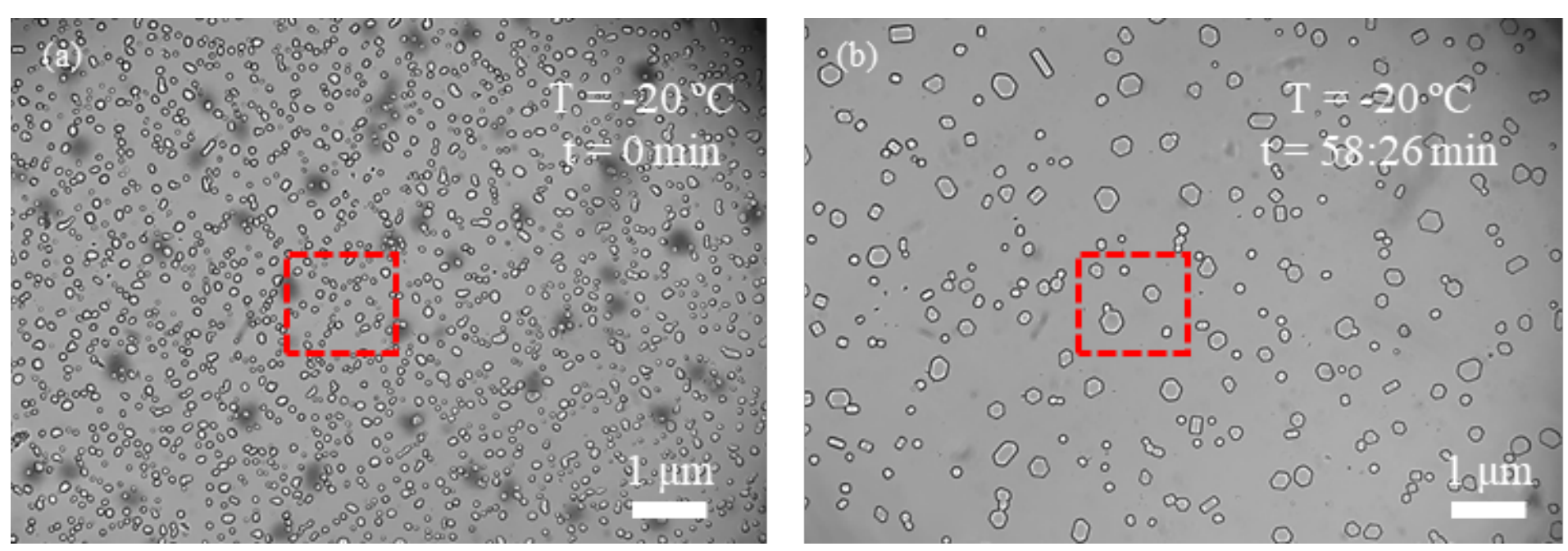

**Figure SI4.** Ice crystals morphology during recrystallization at -20 °C. (a) Initial state at the beginning of the isothermal hold; (b) same field of view after 58 minutes and 26 seconds. Images were acquired with a 10X objective. The red dotted square highlights the region magnified in subsequent images captured using a 50X objective.